\let\mymarginpar\marginpar

\documentclass[10pt, conference, compsoconf]{IEEEtran}
\usepackage[table]{xcolor}
\usepackage{cite}
\usepackage{multirow}
\usepackage{hyperref}
\usepackage{array}
\usepackage{tikz}
\usetikzlibrary{decorations.pathmorphing}
\usepackage{listings}
\usepackage{xcolor}
\usepackage{amssymb}
\usepackage{amsmath}
\usepackage{amsfonts}
\usepackage{xspace}
\usepackage{url}
\usepackage{underscore}
\usepackage{footnote}

\usepackage{dsfont}
\usepackage{textcomp}
\usepackage{colortbl}

\definecolor{tableShade}{HTML}{F1F5FA}   
\definecolor{tableShade2}{HTML}{ECF3FE} 

\newif\ifdraft\draftfalse

\ifdraft
	\newcommand\remark[1]{%
		\mymarginpar{\raggedright\hbadness=10000\tiny\it #1\par}}
\else
	\newcommand\remark[1]	{}
\fi

\ifdraft
\fi

\def\svn$I#1: #2 #3 #4 #5 #6 ${\def\svnrev{#3}\def\svndate{#4 #5}\def\svndoc{#2}}
\svn$Id: x.x 89 2010-08-16 11:01:25Z x $

\newcommand{\opool}{\mathcal{O}}
\newcommand{\ts}{\mathcal{T}}

\newcommand{\pre}[1]{pre(#1)}
\newcommand{\pwpre}[1]{\mathbf{require}(#1)}
\newcommand{\post}[1]{post(#1)}

\newcommand{\reach}[1]{\rho(#1)}
\newcommand{\tnew}{t^{\mathrm{new}}}
\newcommand{\reduct}[2]{\clubsuit_{#1}^{#2}}
\newcommand{\reductpost}[2]{{\spadesuit_{#1}^{#2}}}

\begin{document}
\title{Stateful Testing: Finding More Errors \\ in Code and Contracts}

\newcommand{\ETH}{${}^*$\xspace}
\newcommand{\UDS}{${}^\dag$\xspace}
\newcommand{\smallemail}[1]{ {\scriptsize $\langle$#1$\rangle$}}


\author{\IEEEauthorblockN{Yi~Wei $\cdot$ Hannes~Roth $\cdot$ Carlo~A.~Furia $\cdot$ Yu~Pei $\cdot$ Alexander~Horton $\cdot$ Michael~Steindorfer $\cdot$ Martin Nordio $\cdot$ Bertrand~Meyer}
\IEEEauthorblockA{Chair of Software Engineering, ETH Zurich, Switzerland\\
\{yi.wei, carlo.furia, yu.pei, martin.nordio, bertrand.meyer\}@inf.ethz.ch\ \ \{haroth, ahorton, msteindorfer\}@student.ethz.ch}}


\newcommand{\autoinfer}{AutoInfer\xspace}  
\newcommand{\autotest}{AutoTest\xspace}  
\newcommand{\figref}[1]{Figure~\ref{#1}}  
\newcommand{\secref}[1]{Section~\ref{#1}} 
\newcommand{\lstref}[1]{Listing~\ref{#1}} 
\newcommand{\tblref}[1]{Table~\ref{#1}}   
\newcommand{\onlinepackage}{\url{http://se.inf.ethz.ch/research/autotest}}
\newcommand{\columnname}[1]{\textit{#1} column\xspace} 
\newcommand{\newbug}{63\%\xspace}
\newcommand{\prtime}{2\%\xspace}
\newcommand{\pr}{\textit{pr-}strategy\xspace}
\newcommand{\hours}{520\xspace}
\newcounter{copyrightbox}
\newcommand{\binding}[1]{\textit{Bindings{#1}}}

\maketitle

\begin{abstract}
Automated random testing has shown to be an effective approach to finding faults but still faces a major unsolved issue: how to generate test inputs diverse enough to find many faults and find them quickly. Stateful testing, the automated testing technique introduced in this article, generates new test cases that improve an existing test suite. The generated test cases are designed to  violate the dynamically inferred contracts (invariants) characterizing the existing test suite. As a consequence, they are in a good position to detect new errors, and also to improve the accuracy of the inferred contracts by discovering those that are unsound. 

Experiments on 13 data structure classes totalling over 28,000 lines of code demonstrate the effectiveness of stateful testing in improving over the results of long sessions of random testing: stateful testing found 68.4\% new faults and improved the accuracy of automatically inferred contracts to over 99\%, with just a 7\% time overhead.


\end{abstract}



\pagenumbering{arabic}

\section{Introduction}
Drawing inputs at random may sound like a desultory approach to testing, since it ignores any information about the structure of the system under test.
This intuition, however, turns out to be largely flawed: there is now a compelling amount of evidence---both empirical~\cite{CPOLM11} and analytical~\cite{Simula.se.735}---showing that random testing is a quite effective testing technique that can uncover many subtle errors in real programs.

When the tested software is equipped with \emph{contracts} (pre and postconditions) random testing even becomes a completely \emph{automated} technique: preconditions help select valid inputs and postconditions provide oracles to check if a test case exposes unexpected behavior that does not conform to specification.
The applications of random input generation are not limited to testing but extend to other software dynamic analysis techniques, such as inference of contracts~\cite{WFKM11-ICSE11} (improving and completing those written by programmers) and even automated program correction~\cite{WPFSBMZ10-ISSTA10}.

Constructing random inputs is straightforward for primitive types, such as integers and characters, where it boils down to drawing pseudo-random numbers.
Constructing random objects of arbitrary classes is more involved, because objects can only be created and modified using a class' routines (methods).
To approach this problem, random input generation algorithms for object-oriented languages maintain an \emph{object pool}, which stores all objects randomly generated during the current testing session.
The pool is populated either with fresh objects, built from scratch by creation procedures (constructors), or with objects returned by random routine calls on objects of appropriate type, randomly drawn from the pool.
Routines and creation procedures with arguments are handled by recursively drawing from the pool conforming objects to be used as arguments.
A \emph{test case} is the combination of any target object in the pool with a routine applied to it.

Random testing sessions must last several hours to maximize error-finding effectiveness~\cite{Simula.se.735,CPOLM11}.
A drawback of this necessity is that the object pool grows to contain a large number of objects, even when duplicates are pruned. 
Therefore, the probability of generating at random test cases that would expose new bugs significantly decreases over time: the objects needed to generate the ``missing'' test cases may already be in the object pool, but they are unlikely to be drawn at random because they constitute only a small fraction of the whole pool.

This paper presents \emph{stateful testing}, a dynamic analysis technique that builds on top of random testing and magnifies its effectiveness.
Stateful testing takes over where random testing gives up: after long sessions of random test case generation, the number of faults found reaches a plateau or grows sluggishly, and the object pool contains thousands of objects.
At this point, stateful testing populates a database with the content of the pool stored as serialized objects; the database is searchable for objects that satisfy given predicates.
For example, we can look up an object \lstinline|n| of class \mbox{\lstinline|INTEGER|} such that \lstinline|n > 0|, or an object \lstinline|s| of class \mbox{\lstinline|SET|} that satisfies \linebreak\lstinline|not s.is_empty| (that is, the set contains at least one element).

After populating the database, stateful testing runs dynamic contract inference~\cite{WFKM11-ICSE11} on all \emph{passing test cases} generated during random testing; the result of this step is a collection of pre and postcondition clauses that summarize the properties of the test cases.
Dynamic contract inference characterizes the passing test cases with pre and postconditions based on \emph{templates}, which capture recurring usage patterns that lend themselves to ``meaningful'' generalization.
For object-oriented programs, the set of public \emph{queries} (functions) of a class often provides a valuable collection of predicates to be combined in templates; \lstinline|is_empty| in the example above is a public query that often appears in contracts (inferred and programmer-written).
Since the inference is based on a finite number of observations and on heuristics in the form of templates, some of the inferred contracts can be \emph{unsound}: they merely are a reflection of the test cases that have been exercised.

Stateful testing combines the information stored in the database of objects and the inferred contracts, with the goal of mutually enhancing the test suite and the contracts, along the lines of Xie and Notkin's proposal~\cite{XieN03}.
Stateful testing proceeds by systematically searching the database for objects that \emph{violate} some of the inferred contracts and therefore enable the creation of \emph{new} test cases.
A new test case that executes successfully shows that an inferred contract can be violated without compromising execution, hence the contract is unsound and should be discarded.
A new test case that triggers a failure exposes an faults overlooked in the previous testing session, corresponding to an input never tried before.
Either way, the new test cases improve over the previous testing session by reaching out regions of the object space previously unexplored.
Take, for example, a routine \lstinline|wipe_out| \linebreak of class \lstinline|SET|, which removes all the elements contained in the set.
If \mbox{\lstinline|wipe_out|} has always been called on empty sets, dynamic contract inference suggests the precondition \mbox{\lstinline|is_empty|.}
Then, select an object \lstinline|s| that violates the precondition, that is such that \lstinline|not s.is_empty|.
If the call \lstinline|s.wipe_out| succeeds, it shows that the inferred precondition \lstinline|s.is_empty| is unsound and should be removed.
If the call triggers a failure, it exposes a fault in the routine's implementation, which does not handle correctly sets that are not already empty.

We implemented stateful testing within our AutoTest~\cite{bertrand2009} \linebreak framework for random testing of object-oriented Eiffel applications; the implementation is integrated in EVE~\cite{EVE}, the freely available research branch of the EiffelStudio development environment.
In an extensive set of experiments described in the paper, we applied stateful testing to the historical data generated by running AutoTest for 520 hours on 13 classes from the EiffelBase~\cite{EiffelBase} and Gobo~\cite{Gobo} data structure collections. 
Both libraries have a long development history and are widely used in the Eiffel community. 
AutoTest generated 149,293 distinct test cases, exposed 95 faults in the libraries, and inferred hundreds of new contracts.
We applied stateful testing for 36 hours on this massive data set.
In this relatively limited amount of time, stateful testing exposed 65 new faults (68.4\% improvement) and invalidated 39.3\% of the inferred contracts; manual inspection reveals that almost all the retained contracts are sound.
These figures are promising and demonstrate that stateful testing is an effective technique to boost the effectiveness of random testing and dynamic analysis. 

The rest of the paper is organized as follows: Section~\ref{sec:examples-overview} gives an overview of stateful testing with a few examples; Section~\ref{sec:how-it-works} describes the details of the technique; Section~\ref{sec:semantic} outlines the design of the relational database used to store the results of the initial dynamic analysis; Section~\ref{sec:evaluation} reports the experimental evaluation of stateful testing; Section~\ref{sec:threats} discusses limitations and future work; Section~\ref{sec:related} presents related work; Section~\ref{sec:conclusion} concludes.

\section{Examples} \label{sec:examples-overview}
This section presents three detailed examples that demonstrate the applicability of stateful testing; the examples are from the libraries EiffelBase and Gobo.


\subsection{Unsound preconditions} \label{sec:pre-reduction}
The first example shows how stateful testing can generate tests with a better coverage and detect unsound preconditions.
Class \lstinline|TWO_WAY_SORTED_SET| is the standard Eiffel implementation of sets with ordered elements.
The class includes a public routine 
\begin{lstlisting}[numbers=none]
        merge (other: TWO_WAY_SORTED_SET) 
\end{lstlisting}
which inserts all elements of \lstinline|other| into the \lstinline|Current| set (\emph{this} in Java or C\#).
After running for 40 hours, AutoTest reports a dynamically inferred precondition for \lstinline|merge|:
\begin{lstlisting}[numbers=none]
             pre_1: Current.disjoint (other) ,
\end{lstlisting}
indicating that it has only been called on disjoint sets: \linebreak \lstinline|Current $\cap$ other = $\emptyset$|, hence the functionality of \lstinline|merge| has not been tested thoroughly.

Stateful testing takes over from this situation and tries to generate new test cases that cover the deficiency.
To this end, it looks up the database---filled with data from hours of random testing---for objects of suitable type that \emph{violate} \lstinline|pre_1|; namely, it searches for two objects \lstinline|o1, o2| such that:
\begin{lstlisting}[numbers=none]
        (1)  o1.type = TWO_WAY_SORTED_SET ,
        (2)  o2.type = TWO_WAY_SORTED_SET ,
        (3)  not o1.disjoint (o2) .
\end{lstlisting}
Even if AutoTest never drew such objects during the 40-hour session, there are several pairs satisfying the three constraints (1--3) in the database.
For every such pair of objects, stateful testing generates the new test case \lstinline|o1.merge (o2)|.


Executing the new test cases improves the coverage of routine \lstinline|merge|; 
it also reveals that the inferred precondition \lstinline|pre_1| is unsound and must be \emph{reduced}, hence removing an error in the inferred contracts.
In our experiments, the new test cases did not expose any faults in the implementation of \lstinline|merge|.


\subsection{Unsound postconditions} \label{sec:post-reduction}
The second example shows how stateful testing can detect unsound dynamically inferred \emph{postconditions}.
Routine \lstinline|merge_left (other: LINKED_LIST)| in class \lstinline|LINKED_LIST| merges the content of  \lstinline|other| into the \mbox{\lstinline|Current|} list.
Extensive dynamic analysis reports, among others, the following postcondition for \lstinline|merge_left|:
\begin{lstlisting}[numbers=none]
     post_2: old Current.is_equal (other)
                     implies Current.is_empty .
\end{lstlisting}
That is, whenever \lstinline|Current| and \lstinline|other| contain the same elements (they are \emph{equal}), they are actually empty lists.
\lstinline|post_2| is unsound, as it merely reflects the fact that the test suite never ran \lstinline|merge_left| on lists that are equal but not empty.

Stateful testing targets the antecedent in the implication \lstinline|post_2|, which refers to the state \emph{before} executing \mbox{\lstinline|merge_left|} by means of the \lstinline|old| notation.
The structure of the postcondition suggests to exercise the routine on objects \lstinline|o1, o2| where \mbox{\lstinline|old o1.is_equal(o2)|} is the case, but \lstinline|not o1.is_empty|, with the hope of showing that \lstinline|post_2|'s consequent does not hold after the call.
Stateful testing creates a new test case \mbox{\lstinline|o1.merge_left (o2)|} for every pair of objects in the database that satisfy the criteria.
Since \lstinline|merge_left| does not remove any element from the target \lstinline|o1|, \lstinline|not o1.is_empty| still holds after executing the test cases, thus invalidating \lstinline|post_2| and increasing the coverage of \lstinline|merge_left|.


\subsection{Constructing new objects} \label{sec:unexplored-states}
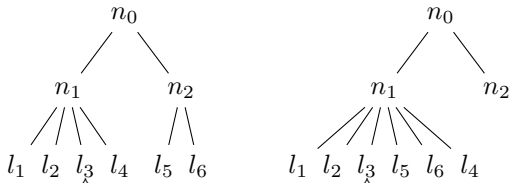
\begin{figure}[!h]
\begin{center}
\begin{tikzpicture} [level 2/.style={sibling distance=1.3em}, level distance=1cm]
\draw (0,0) node (root) {$n_0$}
  child {
   node (tree 1) {$n_1$}
    child { node {$l_1$}}
    child { node  {$l_2$}}
    child { node (marked) {$l_3$} node[below=1pt]{$\Uparrow$}}
    child { node {$l_4$}}}
  child {
    node (tree 2) {$n_2$}
      child { node {$l_5$}} 
      child { node  {$l_6$}}};
\draw (4.2,0) node (root 2) {$n_0$}
  child {
   node (tree 3) {$n_1$}
    child { node {$l_1$}}
    child { node  {$l_2$}}
    child { node (marked 2) {$l_3$} node[below=1pt]{$\Uparrow$}}
    child { node {$l_5$}} 
    child { node  {$l_6$}}
    child { node {$l_4$}}}
  child {
    node (tree 2) {$n_2$}};
\end{tikzpicture}
\end{center}
\vspace{-2em}
\caption{Calling \mbox{$n_1.\!\!$\lstinline|merge_tree_after| $(n_2)$} on the left tree results in the tree shown on the right.}
\label{fig:trees}
\end{figure}

%
The third example shows how stateful testing can generate new objects by mutating other objects serialized in the database.
The example targets class \lstinline|TWO_WAY_TREE|, an implementation of trees with arbitrary number of branch\-es at each level.
An object of type \lstinline|TWO_WAY_TREE| encapsulates a tree's node; each node includes a list of references to its \emph{children}---empty if the node is a leaf---and a \emph{cursor}.
The cursor is an iterator over the list of children, pointing to an element in the list or being \emph{off} the list.
Given two nodes $n_1, n_2$, we can merge $n_2$'s children into $n_1$'s by calling \mbox{\lstinline|$n_1$.merge_tree_after ($n_2$)|}: $n_2$'s list merges into $n_1$'s after the position marked by $n_1$'s cursor, as shown in Figure~\ref{fig:trees} where an arrow $\Uparrow$ marks the position of the cursor, when it is not \emph{off}.
The position ``after the cursor'' is not defined if the cursor is \emph{off}; developers wrote a precondition (\lstinline|require| clause) to \lstinline|merge_tree_after| to enforce this constraint on the input:
\begin{lstlisting}[numbers=none]
      merge_tree_after (other: TWO_WAY_TREE)
          require not off
\end{lstlisting}
where \mbox{\lstinline|off|} is a Boolean query that holds when the cursor of the \lstinline|Current| node is \emph{off} (such as for nodes $n_0$ and $n_2$ in Figure~\ref{fig:trees}).

Dynamic analysis with AutoTest reports the dynamically inferred precondition for \lstinline|merge_tree_after|:
\begin{lstlisting}[numbers=none]
          pre_3: not Current.is_sibling (other) .
\end{lstlisting}
\lstinline|pre_3| reveals that \lstinline|merge_tree_after| has never been tested with \emph{sibling} nodes, that is nodes at the same level of the tree (e.g., $n_1$ and $n_2$ in Figure~\ref{fig:trees}).
Correspondingly, stateful testing looks up the database for objects violating \mbox{\lstinline|pre_3|,} suitable to generate new test cases: two objects \lstinline|o1, o2| of type \lstinline|TWO_WAY_TREE| that satisfy \lstinline|o1.is_sibling (o2)| and \linebreak \lstinline|not o1.off|---the latter constraint is \lstinline|merge_tree_after|'s \linebreak programmer-written precondition.

Unfortunately, no pair of objects in the database satisfies all these constraints: there are several trees with sibling nodes, but all of them have their cursor \emph{off}, hence \mbox{\lstinline|merge_tree_after|} cannot be applied.
In such situations, \linebreak stateful testing selects available objects that satisfy \emph{some} of the requirements and searches for \emph{routines} that can mutate the object state to satisfy the missing requirements.
The database also includes information on the behavior of routines, collected during dynamic analysis. 

In the running example, stateful testing searches for a routine of class \lstinline|TWO_WAY_TREE| that can change a node where \lstinline|off| is \lstinline|False| to one where it is \lstinline|True|.
Routine \mbox{\lstinline|start|} moves the cursor to the first child node (if the child list is not empty), hence it satisfies the search criteria.
With this routine, stateful testing generates a new test case for \mbox{\lstinline|merge_tree_after|} as follows. It selects two serialized objects \mbox{\lstinline|o1, o2|}$\,$ of type \mbox{\lstinline|TWO_WAY_TREE|} that are siblings; the test case consists of two consecutive calls:
\begin{lstlisting}[numbers=none]
          o1.start  ;  o1.merge_tree_after (o2) .
\end{lstlisting}

In our experiments, this new test case triggered a failure, showing that \lstinline|merge_tree_after| does not work correctly on sibling nodes.
This fault went undetected in the random testing session, but stateful testing readily exposed it.

\section{How Stateful Testing Works} \label{sec:how-it-works}
This section starts with an overview of how stateful testing works (Section~\ref{sec:overview}), and then describes the details of the technique: what are the products of random testing (Section~\ref{sec:testing}), how stateful testing processes and organizes them (Section~\ref{sec:ot-db}), the role of dynamically inferred contracts (Section~\ref{sec:inference}), and their reduction to produce new test suites (Section~\ref{sec:reduction}).

\subsection{Overview} \label{sec:overview}
\begin{figure*}[!tb]
    \centering
    \includegraphics*[width=17.5cm]{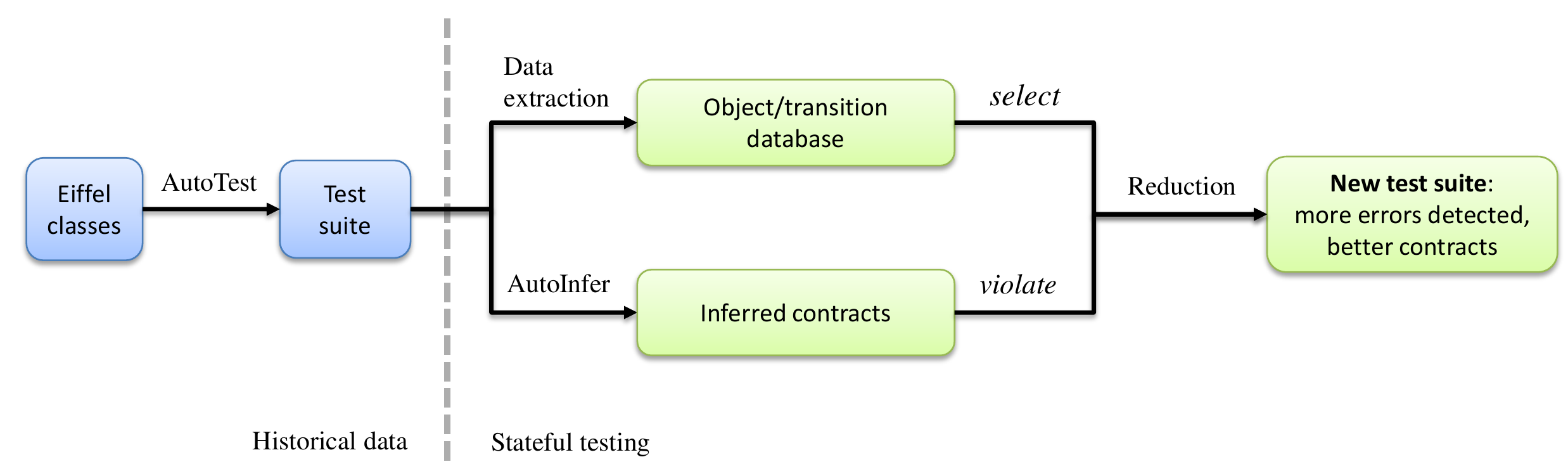}
    \caption{Overview of how stateful testing works.}
\label{fig:overview}
\end{figure*}
\figref{fig:overview} provides a bird's eye view of how stateful testing works.
Stateful testing is a fully automated 
technique that produces new test cases from an existing test suite: %
\begin{enumerate}
\item Running AutoTest, the automatic random testing \linebreak framework for Eiffel, for several hours produces a large pool of \emph{objects}, and a \emph{test suite} based on those objects.
\item Stateful testing selects and extracts information from the object pool and the test suite and stores it in a relational database: the \emph{object/transition database}.
\item AutoInfer, the dynamic contract inference component of AutoTest, summarizes the behavior of the test cases in the test suite in the form of \emph{dynamically inferred contracts}.
\item The \emph{reduction} phase extracts objects from the database that violate some of the inferred contracts. The extracted objects support the generation of a \emph{new test suite}, which exercises the classes under tests differently than in the original test suite. 
\item Executing the new test suite can uncover \emph{new faults} in the code under test, and reveal which of the inferred \emph{contracts} are \emph{incorrect} and should be discarded.  
\end{enumerate}

\subsection{Preliminaries: test cases and objects} \label{sec:testing}
A \emph{test case} $t$ is the call of a routine $r$ on a target object $a_0$ with actual arguments $a_1, \ldots, a_m$ that returns an object $b$, denoted as:
$$t \quad=\quad a_0.r\,(a_1, \ldots, a_m):b\,.$$
If $r$ is a \emph{command} (procedure), which does not return any value, replace $b$ with the dummy object $\epsilon$; if $r$ is a \emph{creation procedure} (constructor), which returns a fresh object, replace the target $a_0$ with $\epsilon$. 

Contracts are annotations using the same syntax as programming language Boolean expressions; they specify the behavior of routines through \emph{preconditions} and \emph{postconditions}.
The precondition of a routine $r$ is a predicate that $r$'s target and arguments satisfy before the call; for example, \lstinline|pre_1| in Section~\ref{sec:pre-reduction} declares that the \lstinline|Current| list (i.e., the target) and the \lstinline|other| list (i.e., the argument) are disjoint, for every call of \lstinline|merge|.
The postcondition of a routine $r$ is a predicate over $r$'s result (if any), as well as $r$'s target and arguments; postconditions can refer to targets and arguments both in the post-state (i.e., after the call) and in the pre-state (i.e., before the call, with the \lstinline|old| keyword).
For example, \lstinline|post_2| in Section~\ref{sec:post-reduction} specifies that, if the target list and the \lstinline|other| list contained the same elements before a call to \lstinline|merge|, then the target is empty after the call.
In Eiffel, programmers can annotate routines with pre (\lstinline|require| clause) and postconditions (\lstinline|ensure| clause); stateful testing includes a contract inference phase that supplements the contracts written by programmers with inferred contracts.

Contracts provide a criterion to determine if a test case is \emph{passing} or \emph{failing} completely automatically.
A routine's test case $a_0.r\,(a_1, \ldots, a_m): b$ is \emph{valid} if its target and arguments $a_0, a_1, \ldots, a_m$ satisfy $r$'s precondition, and is \emph{invalid} otherwise.
Executing a valid test case $t$ changes the target and arguments into the \emph{post-state} $a_0', a_1', \ldots, a_m'$, denoted
$$t \leadsto \langle a_0', a_1', \ldots, a_m' \rangle\,.$$
$t$ is \emph{passing} if executing the test case triggers no exceptions, and the post-state $\langle a_0', a_1', \ldots, a_m' \rangle$, the pre-state $\langle a_0, a_1, \ldots,\linebreak a_m \rangle$ and the returned object $b$ satisfy $r$'s postcondition; otherwise, $t$ is \emph{failing}.

Stateful testing builds upon an existing test suite that exercises a set of classes.
A test suite is a collection $\ts = \{ t_1, t_2, \ldots \}$ of test cases; it induces the set $\opool = \{o_1, o_2, \ldots \}$ of all objects mentioned in $\ts$'s test cases or in the post-state of passing test cases; $\opool$ is the \emph{object pool}.
Stateful testing works independently of how the object pool $\opool$ and the test suite $\ts$ are generated.
Its implementation in the AutoTest framework, however, generates them completely automatically from a set of Eiffel classes with random testing.

\begin{lstlisting}[numbers=none,numberstyle=\tiny,float=!tb,label=lst:list-ex={(*}{*)},caption=Routines of class \lstinline|LIST| with contracts.]
  make: LIST            -- Create an empty list
      ensure Result.is_empty

  wipe_out              -- Remove all elements
      ensure is_empty

  extend (v: ANY)       -- Add `v' to the end
      ensure has (v)

  append (other: LIST)  -- Append `other' to the end
      require other /= Void

  has (v: ANY): BOOLEAN -- Does the list include `v'?

  is_empty: BOOLEAN     -- Is the list empty?
\end{lstlisting}
\textbf{Example.} The class \lstinline|LIST| implements dynamic lists; it is modeled after real Eiffel classes, but is simplified for clarity.
Listing~\ref{lst:list-ex} shows the signatures of \lstinline|LIST|'s routines with programmer-written contracts. 
Consider the test suite $T$:
\begin{lstlisting}[numbers=none]
      $t_1$:   $\epsilon$.make : $l_1 \leadsto \langle \epsilon \rangle$
      $t_2$:   $l_1$.wipe_out : $\epsilon \leadsto \langle l_1\rangle $
      $t_3$:   $l_1$.append ($l_1$): $\epsilon \leadsto \langle l_1, l_1 \rangle$
      $t_4$:   $l_1$.extend ($l_1$): $\epsilon \leadsto \langle l_2, l_2 \rangle$
      $t_5$:   $l_1$.is_empty: $b_3 \leadsto \langle l_1 \rangle$
      $t_6$:   $\epsilon$.make : $l_4 \leadsto \langle \epsilon \rangle$
\end{lstlisting}
where all test cases are passing.
For simplicity, we do not introduce new duplicate objects in $T$ when they are unchanged in the post-state with respect to the pre-state; for example, $t_4$ denotes a call to \lstinline|extend| with $l_1$ as target and argument, and $l_2$ is the name given to the list after extending it, whereas $t_5$ does not change the target $l_1$ which is then repeated in the post-state.
$T$ induces the object pool $O = \{l_1, l_2, b_3, l_4\}$, where $l_1, l_4$ are empty lists, $l_2$ has one element (a reference to $l_2$ itself), and $b_3$ is the Boolean \lstinline|True|.

\subsection{Object/transition database} \label{sec:ot-db}
The object/transition database contains detailed information about the objects in the object pool $\opool$.
Section~\ref{sec:semantic} details how the database is implemented with relational database technology; the current section describes how stateful testing selects and extracts the information to store in the database.
%

\textbf{Abstract object states and transitions.}
The ob\-ject/transition database stores all objects in the pool $\opool$ in serialized form.
On top of the serialized objects, the database stores their \emph{abstract state}, expressed in terms of the public \emph{queries} (functions) of the objects.
In the running example, class \lstinline|LIST| has two public queries: \lstinline|is_empty| and \lstinline|has|.
We would like to have information as extensive as possible in the database: for every combination of objects in the pool, evaluate every public query that is applicable.
This is clearly unfeasible for object pools of non-trivial size, hence stateful testing uses a heuristic based on the usage of objects in the test suite $\ts$.
For an object $o \in \opool$, consider the set $\reach{o}$ of objects \emph{reachable} by recursively following references among $o$'s attributes, and including $o$ itself; because of how the object pool is defined, $\reach{o}$ is a subset of $\opool$.
Extend the notation to objects reachable from a set of objects: $\reach{O} = \bigcup_{o \in O} \reach{o}$.
For every test case $t = a_0.r\,(a_1, \ldots, a_m): b \leadsto \langle a_0', \ldots, a_m' \rangle$, the database stores \emph{all the applicable public queries} $q$:
\begin{gather}
\alpha_0.q\,(\alpha_1, \ldots, \alpha_n): \beta \label{ev:pre} \\
\omega_0.q\,(\omega_1, \ldots, \omega_n): \psi \label{ev:post}
\end{gather}
where $\alpha_0, \alpha_1, \ldots, \alpha_n$ range over the set $\reach{a_0, a_1, \ldots, a_m}$ of object reachable in $t$'s pre-state, and $\omega_0, \omega_1, \ldots, \omega_n$ range over the set $\reach{b, a_0', a_1', \ldots, a_m'}$ of object reachable in $t$'s post-state.
Precisely, for every call of the form (\ref{ev:pre}) or (\ref{ev:post}), the database adds the objects $\beta, \psi$ in serialized form and includes a tuple with $q$'s signature, and references to the serialized objects $\alpha_0, \ldots, \alpha_n, \omega_0, \ldots, \omega_n, \beta, \psi$ stored in the database.
The current tool implementation supports queries of generic return type; for simplicity, the presentation in this paper only considers queries that return Boolean values.

The database also stores information about \emph{transitions}: each transition associates the routine $r$ with several pairs of query evaluations; the first element of the pair evaluates a query in the pre-state (\ref{ev:pre}), and the second evaluates it in the post-state (\ref{ev:post}).
Then, the transition represents the fact that calling $r$ when the pre-state holds can drive the object to the post-state.

Continuing the running example (Listing~\ref{lst:list-ex}), the test cases $t_1, t_2, t_3$ only mention the list object $l_1$, which produces the queries \lstinline|$l_1$.is_empty: True| and \lstinline|$l_1$.has ($l_1$): False|.
$t_4$ introduces the object $l_2$, hence the new queries \mbox{$l_2$\lstinline|.is_empty:False|,} \mbox{$l_1$\lstinline|.has| ($l_2$)\lstinline|:False|,} \mbox{$l_2$\lstinline|.has| ($l_1$)\lstinline|:False|,} \mbox{$l_2$\lstinline|.has| ($l_2$)\lstinline|:True|.}
$t_6$ introduces two more queries on $l_4$: \lstinline|$l_4$.is_empty: True| and \mbox{\lstinline|$l_4$.has ($l_4$): False|}.
Finally, $t_4$ induces the only non-trivial transitions from a pre-state where \lstinline|is_empty| evaluates to \lstinline|True| and \lstinline|has| to \lstinline|False|, to a post-state where both queries change their returned value when evaluated on the changed target.

\textbf{Public branch and path conditions.}
To increase the precision of the abstract states stored in the database, stateful testing includes the value of several Boolean expressions extracted from the program text.
For every test case $t$ exercising a routine $r$, collect all the Boolean expressions $e_1, e_2, \ldots$ that appear as \emph{branch conditions} or as \emph{path conditions} in $r$'s control-flow graph, and that only reference \emph{public features} (members) of $r$'s containing class.
The rationale for storing branch and path conditions is that they often offer ``interesting'' partitions of the input states. 
The database stores the evaluations of these expressions for each applicable combination of objects reachable in the pre-state and in the post-state of every test $t$.



\subsection{Dynamic contract inference} \label{sec:inference}
To get a concise characterization of the test suite $\ts$ in terms of class features, stateful testing performs \emph{contract inference} with dynamic techniques. 
The implementation uses AutoInfer~\cite{WFKM11-ICSE11}, the inference component of the AutoTest framework.

Contract inference only considers the passing test cases from the suite $\ts$ and produces, for each routine $r$ exercised in the test suite, a list $\pre{r}$ of preconditions and a list $\post{r}$ of postconditions.
These inferred contracts summarize $r$'s behavior with the test cases in $\ts$: for every passing test $t = a_0.r\,(a_1, \ldots, a_m): b \leadsto \langle a_0', \ldots, a_m' \rangle$ in $\ts$, the arguments $a_1, \ldots, a_m$ and the target satisfy all preconditions in $\pre{r}$, and the result $b$ (if any) and post-state $a_0', a_1', \ldots, a_m'$ satisfy all postconditions in $\post{r}$.

In the running example (Listing~\ref{lst:list-ex}), \lstinline|wipe_out| is always invoked on an empty list, hence \lstinline|is_empty| is an inferred precondition in \mbox{$\pre{\text{\lstinline|wipe_out|}}$;} \lstinline|append| is invoked once on an empty list which is still empty after the call, hence \lstinline|old is_empty implies is_empty| is a postcondition in \mbox{$\post{\text{\lstinline|append|}}$,} and \lstinline|other.is_empty| is a precondition in \mbox{$\pre{\text{\lstinline|append|}}$.}\footnote{Dynamic inference does not really infer contracts based on so few test cases because they are statistically insignificant; the example is only for illustration purposes.}

The inferred contracts are typically different than those programmers write: the former tend to be more detailed and numerous than the latter, especially in the case of postconditions, which programmers neglect but dynamic analysis is effective at reporting~\cite{PolikarpovaCM2009,WFKM11-ICSE11}.
Furthermore, dynamically inferred contracts have no guarantee of being correct: since they are based on a finite number of observations, they may merely be a reflection of a not sufficiently varied test suite, such as the two examples discussed in the previous paragraph.

\subsection{Reduction} \label{sec:reduction}
After building the object/transition database and collecting the inferred contracts, stateful testing generates a new test suite by \emph{precondition reduction}. 
The basic idea is partitioning the input space: a predicate $p$ defines two regions, one where $p$ holds and one where it doesn't; a comprehensive test suite should cover every region, for every combination of ``interesting'' predicates, with at least one test case.
This is clearly unfeasible, because the predicates are too many; 
precondition reduction is a heuristic technique that considers a reduced number of partitions based on the inferred preconditions.

\subsubsection{Precondition reduction} \label{sec:reduct}
The \emph{precondition reduction} of a routine $r$ generates new inputs to test $r$ by trying to invalidate $r$'s inferred preconditions.
Suppose $r$ has $m$ arguments, and let $\pwpre{r}$ denote $r$'s programmer-written preconditions.
Select a dynamically inferred precondition $p$ from the set $\pre{r}$ and build the predicate:
$$\reduct{p}{r}:\quad \neg p \ \wedge\ \pwpre{r}\,.$$
$\reduct{p}{r}$ characterizes objects that satisfy $r$'s programmer-written preconditions but violate the inferred $p$, hence they can be used to test $r$ in a way not covered by the existing test suite.

Stateful testing searches the object/transition database for tuples of objects $\langle o_0, o_1, \ldots, o_m\rangle$ that satisfy $\reduct{p}{r}$ (expressed as a conjunction of elementary expressions).
In the running example (Listing~\ref{lst:list-ex}), \lstinline|wipe_out|'s inferred precondition \lstinline|is_empty| suggests to search for objects of type \lstinline|LIST| satisfying \lstinline|not is_empty| (\lstinline|wipe_out| has no programmer-written precondition); $l_2$ satisfies the search criterion.

For each tuple $\langle o_0, o_1, \ldots, o_m\rangle$ retrieved in the search, stateful testing constructs the new test case $$\tnew = o_0.r\,(o_1, \ldots, o_m)\,.$$
In practice, there is a cut-off on the number of retrieved tuples (if they are too many, only a few are tried) and a time-out on the time spent searching the database (if no tuple is found by the time-out, we move to the next reduction).
If $\tnew$ is passing, then the precondition $p$ is unsound and removed from $\pre{r}$; if $\tnew$ is failing, a fault is found (and $p$ is also unsound).
Since the information stored in the database is incomplete, $\tnew$ may also be invalid, in which case it is simply discarded.
In the running example, the list $l_2$ is not empty and the test case \lstinline|$l_2$.wipe_out| exercises \lstinline|wipe_out| in ways not tested before.

\subsubsection{Using transitions} \label{sec:reduct-transitions}
If the search for objects satisfying $\reduct{p}{r}$ fails, stateful testing tries to retrieve objects satistying a weaker predicate than $\reduct{p}{r}$, and then it searches for a transition that drives the objects to match the desired $\reduct{p}{r}$.
To this end, put $\reduct{p}{r}$ in conjunctive normal form $c_1 \wedge \cdots \wedge c_n$ and select $1 \leq d < n$ clauses to drop; without loss of generality, we drop $c_1, \ldots, c_d$  ($D$) and we keep $c_{d+1}, \ldots, c_n$ ($K$):
$$\reduct{p}{r} \equiv \quad \underbrace{c_1 \wedge \cdots c_d}_{D} \,\wedge\, \underbrace{c_{d+1} \wedge \cdots \wedge c_n}_{K}\,.$$
For any tuple of objects $\langle o_0, \ldots, o_m \rangle$ satisfying $K$, search the object/transition database for \emph{transitions} that can transform a tuple $\!\langle o_0, \ldots, o_m \rangle$ satisfying $\neg D$ into a tuple $\langle o_0', \ldots, \linebreak o_m' \rangle$ satisfying $D$.
Every such transition consists of a routine $s$ and a mapping $\mu: [0..n] \rightarrow [0..m]$, where $s$ has $n \geq 0$ arguments.
$\mu$ binds the objects $o_0, \ldots, o_m$ to $s$'s target and arguments: the $i$-th argument is instantiated with $o_{\mu(i)}$.
For every such transition, construct the new test case $$\tnew = o_{\mu(0)}.s\,(o_{\mu(1)}, \ldots, o_{\mu(n)}) \;;\;  o_0.r\,(o_1, \ldots, o_m)\,,$$
consisting of two consecutive calls.

In the \lstinline|TWO_WAY_TREE| example in Section~\ref{sec:unexplored-states}, the dropped clause $D$ is \lstinline|not Current.off| and the kept clause $K$ is \lstinline|Current.is_sibling (other)|. 
Two objects \lstinline|o1, o2| in the da\-ta\-base satisfy \lstinline|o1.is_sibling (o2)|, and a transition suggests that routine \lstinline|start| can change \lstinline|o1| from \mbox{\lstinline|o1.off|} to \lstinline|not o1.off|.

The search for transitions is heuristic: since the information about transitions in the database is incomplete in general, the routine $s$ may be inapplicable to the objects $\langle o_0, \ldots, o_m \rangle$, or it may not drive them in a state satisfying $\reduct{p}{r}$---for example, because it invalidates $K$ as a side-effect of satisfying $D$.
In practice, the heuristic search is reasonably successful when there are objects whose state is close to satisfying $\reduct{p}{r}$; correspondingly, the current implementation drops at most one clause ($d = 1$), and does not build sequences of transitions with more than two calls.

\subsubsection{Detecting unsound postconditions} \label{sec:reductpost}
Inferred postconditions can be unsound, too, but we cannot directly select objects that violate postconditions, because we do not have direct control over post-states.
Precondition reduction, however, can also help to invalidate inferred postconditions, while testing routines more thoroughly.
Consider an inferred postcondition $q$ in $\post{r}$ in the form:
$$q:\quad \text{\lstinline|old|} (A) \,\Longrightarrow\, C\,.$$
We focus on postconditions in this form, because $q$ naturally expresses many postconditions where a property $C$ of the post-state is a consequence of a property $A$ of the pre-state \mbox{(\lstinline|old|)}.
%
Invalidating the implication $q$ means producing test cases that start in a pre-state where $A$ holds and reach a post-state where $\neg C$ holds.
The existing test suite does not include such test cases, otherwise $P$ would not be a dynamically inferred postcondition.

The inferred \emph{preconditions}, however, help select pre-states that may challenge the validity of $q$. 
%
To this end, consider the set $\pre{r|A}$ of $r$'s dynamically inferred preconditions that hold when $A$ also holds.
Select a $p \in \pre{r|A}$ among these preconditions and build the predicate:
$$\reductpost{p,q}{r}: \quad A \ \wedge\ \reduct{p}{r}\,.$$
Then, select (or build with transitions) objects $\langle o_0, \ldots, o_m \rangle$ that satisfy $\reductpost{p,q}{r}$, and generate the new test case $\tnew$ that calls $r$ on $\langle o_0, \ldots, o_m \rangle$ (as in Section~\ref{sec:reduct}).
If $\tnew$ is valid and passing (with respect to $r$'s programmer-written contracts only) but $C$ is false after executing it, the postcondition $q$ is unsound and is removed from $\post{r}$; if $\tnew$ is failing (again with respect to $r$'s programmer-written contracts, which are always assumed correct), it also shows a fault.

In the example of Listing~\ref{lst:list-ex}, stateful testing targets the inferred postcondition \lstinline|old is_empty implies is_empty| of routine \lstinline|append|, which is in the form $q$: $A$ is \lstinline|is_empty|; for the same routine, \lstinline|other.is_empty| is a precondition inferred when $A$ also holds.
Hence, stateful testing looks for two lists, one empty and one not; $l_1, l_2$ satisfy the criterion and yield the new test case \lstinline|$l_1$.append ($l_2$)|.

\section{Object/transition database}
\label{sec:semantic}
Section~\ref{sec:ot-db} describes what kind of information the \linebreak object/transition database stores; the present section details how the database is implemented with relational technology.

\begin{figure}[!h]
    \centering
    \includegraphics[width=8.3cm]{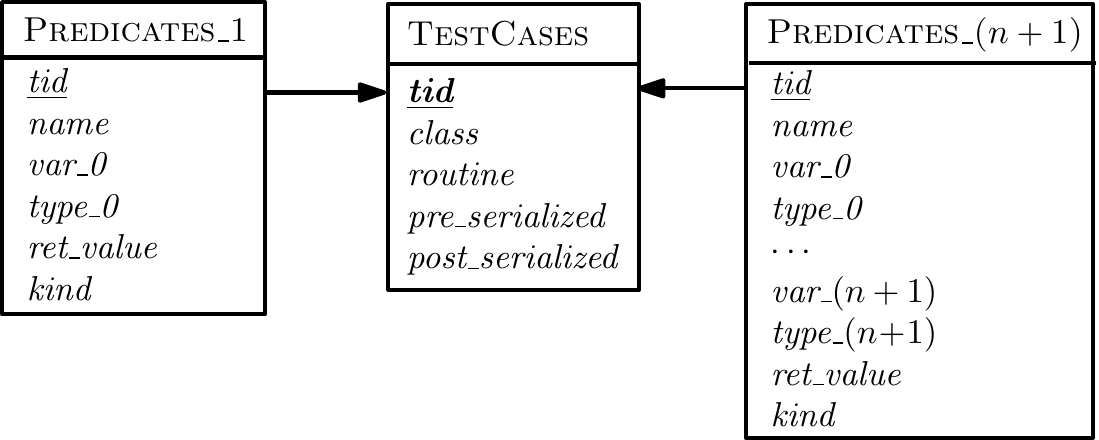}
    \caption{Relational schema of the object/transition database.}
\label{fig:schema}
\end{figure}

\subsection{Relational schema}
Figure~\ref{fig:schema} shows the most significant parts of the \linebreak object/transition database's relational schema.
The da\-ta\-base is centered around the test cases in the test suite $\ts$: for each test case $a_0.r\,(a_1, \ldots, a_m): b \leadsto \langle a_0', \ldots, a_m'\rangle$ table \textsc{TestCases} stores: (1) a unique identifier (attribute \emph{tid}), (2) $r$'s class (\emph{class}), (3) $r$'s name (\emph{routine}), (4) the list $a_0, a_1, \ldots, a_m, a_{m+1}, \ldots, a_n$ of all serialized objects in the pre-state followed by all objects reachable from them (\emph{pre_serialized}), (5) the list $a_0', a_1', \ldots, a_m', b, a_{m+1}', \ldots, a_{n'}'$ of all serialized objects in the post-state followed by all objects reachable from them (\emph{post_serialized}).
We do not discuss the straightforward details of how lists of serialized objects are encoded as sequences of characters with separators.

The other tables \textsc{Predicates_1}, \textsc{Predicates_2}, \ldots, \linebreak \textsc{Predicates_9} store information about the abstract state of objects, in the form of predicates over 1, 2, \ldots, 9 objects.
Consider an atomic Boolean predicate $q$ evaluated over $n+1$ objects, in the form $o_0.q\,(o_1, \ldots, o_n): v$, that holds for the pre-state of a test case with identifier $x$. \linebreak
Table \textsc{Predicates_$(n+1)$} stores an entry with: (1) a reference to the test case $x$ (attribute \emph{tid}), (2) the normalized textual form of the predicate, obtained by replacing every reference to objects with `\$' placeholders as in $\$.q\,(\$, \ldots, \$)$ (attribute \emph{name}), (3) for each object $o_i$, $0 \leq i \leq n+1$, an integer $k_i$ such that the $k_i$-th element of the list in the \emph{pre_state} attribute of the test case with \emph{tid} $x$ contains $o_i$ (attribute \emph{var_i}), (4) for each object $o_i$, $0 \leq i \leq n+1$, its dynamic type (\emph{type_i}), (5) the Boolean value $v$ returned (attribute \emph{ret_value}), (6) the constant \emph{pre} to denote that $q$ is evaluated over the pre-state (attribute \emph{kind}).
For predicates evaluated in the post-state, attribute \emph{kind} stores the constant \emph{post}, and everything else is like for pre-states.

Consider, for example, the test case $t_4$ in the running example (Section~\ref{sec:testing}): $l_1.\text{\lstinline|extend|}(l_1): \epsilon \leadsto \langle l_2, l_2 \rangle$.
Table \textsc{TestCases} stores a tuple $\langle id_4, \text{\lstinline|LIST|}, \text{\lstinline|extend|}, \pi, \Pi  \rangle$, where $id_4$ is the unique identifier, $\pi$ is the list of (serialized) objects in the pre-state: $\pi = [l_1, l_1]$, and $\Pi$ is the list of objects in the post-state: $\Pi = [l_2, l_2]$.
$t_4$ induces, among others, the evaluation of the query $l_1.\text{\lstinline|has|}(l_1)$ in the pre-state, which table \textsc{Predicates_2} stores as the tuple $$\langle id_4, \$.\text{\lstinline|has|}(\$), 0, \text{\lstinline|LIST|}, 0, \text{\lstinline|LIST|}, \text{\lstinline|False|}, \textit{pre} \rangle$$
where the entries $0$ in attributes \emph{var_0} and \emph{var_1} refer to the first element in the $0$-indexed list $\pi$ of serialized objects (i.e., object $l_1$ in serialized form).

\subsection{Quering the database}
The translation of predicates into SQL queries to the \linebreak object/transition database is straightforward:
\begin{itemize}
\item Objects become variables in the \lstinline[language=SQL]|SELECT| clause;
\item These variables are joined with the \textsc{Predicates} and \textsc{TestCases} tables in the \lstinline[language=SQL]|FROM| clause;
\item The \lstinline[language=SQL]|WHERE| clause encodes the constraints on the individual predicates, and SQL Boolean operators map the Boolean connectives in the translated predicate.
\end{itemize}

Let us demonstrate the creation of SQL queries with the example at the end of Section~\ref{sec:reductpost}, where we search for two objects $l_1, l_2$ of type \lstinline|LIST| such that \lstinline|$l_1$.is_empty| and \linebreak\lstinline|not $l_2$.is_empty|.
We create the SQL query in Listing~\ref{lst:sql-query} that searches for such objects in pre-states (the query for post-states is all similar).
The SQL query returns a tuple \mbox{\lstinline[language=SQL]|objs1,|} \mbox{\lstinline[language=SQL]|objs2, idx1, idx2|} such that \lstinline[language=SQL]|objs1, objs2| are collections of serialized objects and \lstinline[language=SQL]|idx1, idx2| are integer indices: the \linebreak \lstinline[language=SQL]|idx1|-th element in collection \lstinline[language=SQL]|objs1| is an empty list, and the \lstinline[language=SQL]|idx2|-th element in collection \lstinline[language=SQL]|objs2| is a non-empty list.

\begin{lstlisting}[language=SQL, numbers=none, mathescape=false, upquote=true,float=!htb,label=lst:sql-query,caption={An SQL query searching for two lists.}]
SELECT
 t1.pre_serialized as objs1, t2.pre_serialized as objs2,
 p1.var_0 as idx1, p2.var_0 as idx2
FROM
 Predicates_1 p1 join TestCases t1 on p1.tid = t1.tid,
 Predicates_1 p2 join TestCases t2 on p2.tid = t2.tid
WHERE
 p1.name = '$.is_empty'                  AND  
 p1.type_0 = 'LIST'                      AND
 p1.ret_value AND p1.kind = 'pre'        AND

 p2.name = '$.is_empty'                  AND
 p2.type_0 = 'LIST'                      AND
 NOT (p2.ret_value) AND p2.kind = 'pre'
\end{lstlisting}

\section{Evaluation}
\label{sec:evaluation}
\begin{table*}[!bt]
\centering \caption{Classes under test and results.\label{tbl:result}}
\rowcolors{2}{tableShade2}{white}
\begin{tabular}{lrrr | rrr@{\ \ } rrr | rrr@{\ \ } rrr}
\hline
\multicolumn{4}{c|}{\textsc{Random Testing}} & \multicolumn{6}{c|}{\textsc{Stateful testing with preconditions}} & \multicolumn{6}{c}{\textsc{Stateful testing with postconditions}} \\
\textsc{Class}              &  \textsc{LOC}            &  \#R    & \#E  &   \#T$_p$   &  \#U$_p$   & \multicolumn{2}{l}{\#V$_p$} & \#E$_p$   &  \#M$_p$      & \#T$_{q}$ &  \#U$_{q}$ & \multicolumn{2}{l}{\#V$_q$} & \#E$_q$ & \#M$_q$ \\
\hline
ARRAY              &  1466           &    65   &  9   &   111       &   23       &   23 & (100\%)              &    2      &   52$'$       &     14    &     1      &    1 & (100\%)              &   0     &   5$'$    \\         
ARRAYED_QUEUE      &  1064           &    40   &  0   &    17       &   13       &   13 & (100\%)              &    0      &    7$'$       &     19    &     0      &    0 & N/A                  &   0     &   9$'$    \\         
ARRAYED_SET        &  2343           &    46   &  9   &    55       &   18       &   18 & (100\%)              &    1      &   25$'$       &    141    &     0      &    0 & N/A                  &   0     &  10$'$    \\         
BOUNDED_QUEUE      &  1130           &    40   &  0   &    20       &   16       &   16 & (100\%)              &    0      &    7$'$       &     22    &     0      &    0 & N/A                  &   0     &   9$'$    \\         
DS_ARRAYED_LIST    &  2760           &   104   &  5   &   178       &  107       &  107 & (100\%)              &    4      &   92$'$       &    170    &    16      &   11 & (69\%)               &   0     & 154$'$    \\         
DS_HASH_SET        &  3074           &    82   &  1   &   279       &  173       &  173 & (100\%)              &    2      &   40$'$       &     51    &     3      &    3 & (100\%)              &   0     &   5$'$    \\         
DS_LINKED_LIST     &  3432           &   100   &  5   &   196       &  120       &  120 & (100\%)              &    2      &  106$'$       &    129    &     1      &    0 & (0\%)                &   1     &  88$'$    \\         
DS_LINKED_STACK    &   934           &    28   &  0   &    39       &   38       &   38 & (100\%)              &    0      &    4$'$       &      4    &     0      &    0 & N/A                  &   0     &   1$'$    \\         
HASH_TABLE         &  2032           &    58   &  1   &   117       &   88       &   87 & (99\%)               &    1      &   16$'$       &     63    &    10      &   10 & (100\%)              &   0     &  30$'$    \\         
LINKED_LIST        &  1998           &    72   &  1   &    53       &   46       &   46 & (100\%)              &    0      &    9$'$       &    149    &    13      &   13 & (100\%)              &   1     &  22$'$    \\         
LINKED_SET         &  2366           &    80   & 13   &    91       &   47       &   47 & (100\%)              &    4      &   33$'$       &    176    &    15      &   15 & (100\%)              &   1     &  28$'$    \\         
TWO_WAY_SORTED_SET &  2866           &    92   & 29   &   221       &  120       &  120 & (100\%)              &   15      &   49$'$       &     25    &     7      &    7 & (100\%)              &   0     &   2$'$    \\         
TWO_WAY_TREE       &  3316           &   107   & 22   &   364       &  203       &  198 & (98\%)               &   26      &   75$'$       &     10    &     3      &    0 & (0\%)                &   5     &   4$'$    \\         
\hline
\rowcolor{white}
\textbf{Total}     & \textbf{28781}           &   \textbf{914}   & \textbf{95}   &   \textbf{1741}     &  \textbf{1012}       &  \textbf{1006} &\textbf{(99.4\%)}       &   \textbf{57}      &  \textbf{515$'$}       &    \textbf{973}    &    \textbf{68}      &   \textbf{60} & \textbf{(88.2\%)}     &   \textbf{8}       &  \textbf{367$'$}   \\
\hline
\end{tabular}
\end{table*}
This section presents the results of an experimental evaluation, summarized in Table~\ref{tbl:result}: the leftmost part of the table contains statistics about random testing, the middle part shows the performance of stateful testing with preconditions and the rightmost part with postconditions.

\subsection{Experimental setup}
The experiments targeted 13 Eiffel classes implementing data structures from the libraries EiffelBase~\cite{EiffelBase} (revision 506) and Gobo~\cite{Gobo} (revision 6665).
Table~\ref{tbl:result} lists the size of each class in lines of code (\textsc{LOC}) and public routines (\textsc{\#R}).

Each session in the preparation of the original test suite with random testing ran on a Linux node with a 2.53 GHz Intel Nehalem quad-core CPU and 8 GB of memory. 
The other experiments (contract inference and stateful testing) ran on an Ubuntu machine with a 1.73 GHz Intel Core i7 CPU and 8 GB of memory.
The average speed of random testing, contract inference, and stateful testing on the two architectures is comparable.

\subsubsection{Random testing}
To generate the original test suite $\ts$---upon which stateful testing builds---AutoTest ran 30 sessions of random testing for each of the 13 classes.
A session lasts 80 minutes and initializes the pseudo-random number generator with a new seed.
The 30 sessions totaled 520 hours of testing and generated a test suite with 149,293 distinct test cases.
The test suite $\ts$ revealed 95 distinct faults\footnote{Two faults are distinct if they violate two different contract clauses.} (column \textsc{\#E} of Table~\ref{tbl:result}).

\subsubsection{Stateful testing running time}
\textbf{Dynamic contract inference.}
AutoInfer processed the test suite $\ts$ for 16 hours and reported 
1741 preconditions and 973 postconditions expressible as implications, shown in column \#T$_p$ and \#T$_q$ in Table~\ref{tbl:result}.
Manual inspection revealed that 1012~(58\%) of the inferred preconditions and 68~(7\%) of the inferred postconditions are unsound.
Columns \#U$_p$ and \#U$_q$ respectively report the number of unsound pre and postconditions for each class.

\textbf{Object/transition database construction.}$\!$
Con\-struct\-ing the object/transition database from $\ts$ took 5 hours. The database contains about 3.5 million objects, 18.4 million predicate evaluations, and 68.8 thousand transitions, and occupies 3.4 GB on disk.

\textbf{Reduction.}
Notice that querying the object/transition database gives predictable results, hence the reduction is deterministic and needs to run only once.
Stateful testing ran for 15 hours trying to violate the inferred pre and postconditions. The times (in minutes) spent on the pre and postconditions in each class are shown in columns \#M$_p$ and \#M$_q$ of Table~\ref{tbl:result}.
In the experiments, every query times out after one minute.

\subsection{Experimental results}

In all, stateful testing discovered 65 new faults in the classes under test, corresponding to a 68.4\% improvement over the number of faults found by random testing, with only
a 7\% time overhead ($36 / 520$ hours). 
Columns \#E$_p$ and \#E$_q$ in Table~\ref{tbl:result} respectively show the number of new faults detected while trying to violate the inferred pre and postconditions in each class.
The performance in terms of number of unsound preconditions and postconditions detected is given below.
\vspace{0.1cm}
\noindent \framebox{\parbox{8.6cm}{\centering \emph{Building upon random testing, stateful testing detected 68.4\% new faults in a fraction of the time.}}}
\subsubsection{Unsound preconditions}
Table~\ref{tbl:structure} gives an account of the most common structures of the inferred preconditions targeted in the experiments.
Stateful testing tried to invalidate the 1741 inferred preconditions for 8.2 hours (i.e., about 18 seconds per precondition), following the technique in Section~\ref{sec:reduct}.
It successfully invalidated 1006 (99.4\%) of the unsound preconditions (column \#V$_p$ of Table~\ref{tbl:result}, which also report the percentages relative to column \#U$_p$), while exposing 57 new faults (column \#E$_p$).

\begin{table}[!h]
\centering \caption{Structure of inferred preconditions.\label{tbl:structure}}
\begin{tabular}{llr}
\hline
\textsc{Structure}                 & \textsc{Example}                    &  \#T    \\
\hline
Reference equality                 & \lstinline|o1 = o2|                 &   154   \\
Object equality                    & \lstinline|o1.is_equal(o2)|         &   329   \\
Voidness check                     & \lstinline|o /= Void|               &     7   \\
Integer equality                   & \lstinline|o.count = 0|             &   377   \\
Boolean query with arguments       & \lstinline|o.has(v)|                &   483   \\
Boolean query without arguments    & \lstinline|o.is_empty|              &   356   \\
Other (complex)                    & \lstinline|o.full or i < l.count|   &    35   \\
\hline
\textbf{Total}                     &                                     & \textbf{1741}  \\
\hline
\end{tabular}
\end{table}

\subsubsection{Unsound postconditions}
Stateful testing tried to invalidate the 973 inferred postconditions in implication form for 6 hours (i.e., about 23 seconds per postcondition), following the technique in Section~\ref{sec:reductpost}.
It successfully invalidated 60 (88.2\%) of the unsound postconditions (column \#V$_q$ of Table~\ref{tbl:result}, which also report the percentages relative to column \#U$_q$), while exposing 8 new faults (column \#E$_q$).

\subsubsection{Undetected unsound contracts}
Stateful testing only failed to detect 6 unsound preconditions (0.6\% of the total) and 8 unsound postconditions (11.8\%).
In all such cases, no serialized objects were in a state violating the contract (or sufficiently close to it), or the predicates provided an abstraction of the object state that was too coarse-grained for the desired objects to be identifiable.

\noindent \framebox{\parbox{8.6cm}{\centering \emph{Stateful testing increased the soundness of inferred \linebreak contracts from 60.2\% to 99.5\%.}}}

\section{Limitations and future work} \label{sec:threats}
Stateful testing, and its current implementation, has some limitations to be addressed in future work.

\begin{itemize}
\item As it is customary in random testing~\cite{SharmaGAFM11}, we have evaluated stateful testing on classes implementing data structures. This was also useful for comparison against out previous experience with AutoTest~\cite{CPOLM11}. Further experiments will target different types of classes.

\item Stateful testing can start from a test suite $\ts$ generated manually or with any technique, all our experiments used test cases automatically generated. 
  Further experiments will determine if the performance of stateful testing is affected by how the original test suite is generated.

\item The current implementation of stateful testing adopts the following heuristics when retrieving objects and transitions from the da\-ta\-base: 
(1) search for objects and consider the first 45 results; (2) if none of the 45 retrieved objects work, search for transitions and call them on the 45 result objects; (3) if none of the transitions work, give up and move to the next reduction.
 This heuristic worked quite well in the experiments, but further experience will determine if it can be improved and how robust the results are with respect to this heuristic.

\item Future experiments will try to iterate the infer/reduce process on the new test suite generated with stateful testing.
  This will challenge the state of the art in dynamic invariant inference and is likely to suggest improvements to the techniques used in the process.
\end{itemize}

\section{Related Work}
\label{sec:related}
Xie and Notkin~\cite{XieN03} first suggested a framework that combines test-case generation and dynamic specification inference with the goal of mutually enhancing their results.

Dallmeier et al.~\cite{DallmeierKMHZ10} implement Xie and Notkin's ideas for typestate specifications (finite-state automata describing abstract object states and transitions), and report an evaluation showing that their technique builds more accurate specification and finds more errors injected in Java applications than traditional dynamic analysis techniques~\cite{DallmeierZM09}.
Stateful testing is based on the same principles---applied to contract specifications---and extends them with the usage of a database to improve the reuse of previous testing sessions and to build new test cases.
Typestates provide object-state abstractions---based on argumentless Boolean queries and simple integer partitioning---that are coarser-grained than the one deployed in the present paper; consequently, building a typestate model by exhaustive exploration is feasible, whereas our more detailed model requires heuristics and an efficient search of serialized objects to be built.
We also provide an implementation and experimental evaluation.

Stateful testing combines diverse techniques of program analysis; the rest of this section summarizes some representative work involving these techniques.
More comprehensive references are available in the bibliography of the cited work.

\subsection{Automated test-case generation}
Automated random testing~\cite{bertrand2009} is now a well-understood technique which, in spite of the simplicity of its underlying ideas, is quite effective and can find subtle bugs~\cite{CPOLM11}.
Arcuri et al.'s analysis of random testing~\cite{Simula.se.735} analytically confirms the experimental results, and suggests that more sophisticated test-case generation techniques are best deployed \emph{after} random testing exhausts its potential.
Stateful testing is indeed applied following random testing sessions, to reuse the objects generated and find new inputs and errors.

Search-based test-case generation refines random testing with goal-driven searches in the space of test cases; McMinn~\cite{McMinn04} and Ali et al.~\cite{AliBHP10} survey the state of the art in search-based techniques. 
Test suite augmentation~\cite{XuFSE2010} uses search-based techniques driven by coverage criteria, with the purpose of adapting a regression test suite to changed code.
Genetic algorithms are a recurring choice to search for test inputs; Tonella~\cite{Tonella04} first suggested the idea, and Andrews et al.~\cite{AndrewsML11} show how to use genetic algorithms to optimize the performance of standard random testing.
Stateful testing is also search-based, but the search takes place among previously generated objects, and it is guided by contracts that characterize an existing test suite to be improved.

Other refinements of random testing combine it with \linebreak white-box techniques such as symbolic execution (e.g., \cite{MajumdarS07}), or leverage the availability of formal specifications in various forms (see Hierons et al.~\cite{HieronsBBCDDGHKKLSVWZ09} for a survey).
Stateful testing also makes extensive usage of specifications in the form of contracts, both inferred and written by programmers.

In previous work~\cite{WeiICST10}, we developed \emph{precondition satisfiaction}, a search strategy that improves the selection of objects to test routines with complex preconditions.
This technique is included in AutoTest, and all the random testing session that preceded stateful testing (in the experiments of Section~\ref{sec:evaluation}) deployed precondition satisfaction.

\subsection{Dynamic specification inference}
Daikon~\cite{ECGN01} pioneered the dynamic inference of specifications and program invariants, and showed that assertions ``guessed'' based on a finite number of runs are often sound with respect to generic executions.
Since the first Daikon release, dynamic inference has been applied to other specification models (e.g., typestates~\cite{Dallmeier.MOB.2006}) and has improved its accuracy (such as in our own AutoInfer~$\!$\cite{WFKM11-ICSE11}).

Gupta and Heidepriem~\cite{GuptaH03} suggest to improve the quality of inferred contracts by using different test suites (i.e., based on code coverage and invariant coverage), and by retaining only the contracts that are inferred with both techniques.
Fraser and Zeller~\cite{fraserICST11} simplify and improve test cases based on mining recurring usage patterns in code bases; the simplified tests are easier to understand and focus on common usage.
Other approaches to improve the quality of inferred contracts combine static and dynamic techniques (e.g., \cite{CsallnerTS2008,TillmannCS06}).

Stateful testing leverages dynamic contract inference techniques, and tries to violate inferred contracts to explore new regions of the input state space; this not only improves the test suite, but it also detects many unsound contracts.
Stateful testing also includes the results of some lightweight static analysis (based on the branching structure of routines) to gather more information about the available test suite.

\subsection{Search-based techniques}
The idea of constructing ``semantic'' databases, with a uniform search interface to retrieve programs~\cite{GrechanikICSE2010}, program elements~\cite{WurschICSE10,Reiss09}, or test cases~\cite{HolzerASE2010} with specific characteristics has recently been deployed, mostly to help developers organize their code and reuse products written by others.

The object capture technique~\cite{ocat10} stores serialized objects, created during program executions, and reuses them as new test inputs to reach uncovered branches. 
Stateful testing also stores serialized objects and reuses them to create new test cases; the object capture framework, however, only supports searching for objects based on their types, whereas stateful testing stores rich information about the objects' abstract states, as well as transitions between states.
Another significant difference is that stateful testing targets the mutual improvement of test suites and inferred contracts, whereas object capture is only concerned with improving branch coverage.

\section{Conclusions}
\label{sec:conclusion}
This paper presented \emph{stateful testing}, a completely automated testing technique which generates new test cases from an existing test suite.
Stateful testing works by trying to \emph{reduce} (i.e., invalidate) the inferred contracts that characterize the existing test suite.
Extensive experiments show that stateful testing is quite effective: it generates tests that uncover new faults and invalidates many of the unsound contracts inferred dynamically from the original test suite.

Stateful testing is part of the automated testing framework AutoTest; the source code of AutoTest---including the basic testing infrastructure, dynamic contract inference, and the stateful testing implementation---detailed experimental results and instructions to reproduce the experiments are available at:
\begin{center}
\onlinepackage
\end{center}

\textbf{Acknowledgments.}
Nadia Polikarpova suggested the idea---and the name---of precondition reduction, and provided valuable comments on a draft of this paper.
This work has been partially funded by the Swiss National Science Foundation under the projects SATS and ASII (SNF~200021-117995 and 200021-134976); it also benefited from funding by the Hasler foundation on related projects. 
The facilities of the Swiss National Supercomputing Centre (CSCS) helped to generate the large-scale test suite used in the experiments.

\bibliographystyle{IEEEtran}
\bibliography{reduct}

\end{document}